\documentclass{rnaastex}

\begin{document}

\title{The Final Integrations of the Caltech Submillimeter Observatory}

\author{Brett A. McGuire}
\altaffiliation{B.A.M. is a Hubble Fellow of the National Radio Astronomy Observatory}
\affiliation{National Radio Astronomy Observatory}
\email{bmcguire@nrao.edu \& paul.carroll@cfa.harvard.edu}

\author{P. Brandon Carroll}
\altaffiliation{P.B.C. is a Simons Foundation Collaboration on the Origins of Life Postdoctoral Fellow}
\affiliation{Harvard-Smithsonian Center for Astrophysics}


\section{} 

\noindent We present the final observations of the Caltech Submillimeter Observatory, taken toward Orion KL, prior to its decommissioning after nearly three decades of operation.  Early in the morning of 8 September 2015, the sun rose over the Caltech Submillimeter Observatory (CSO), heralding the end of its final night of science observations after nearly three decades of impactful science and cutting-edge instrument development.  The weather had broken for the first time in weeks, and despite an optical depth of $\tau_{\mathrm{225~GHz}} \sim 0.25$, the CSO performed admirably.  \\

\begin{figure*}
\centering
\includegraphics[width=\textwidth]{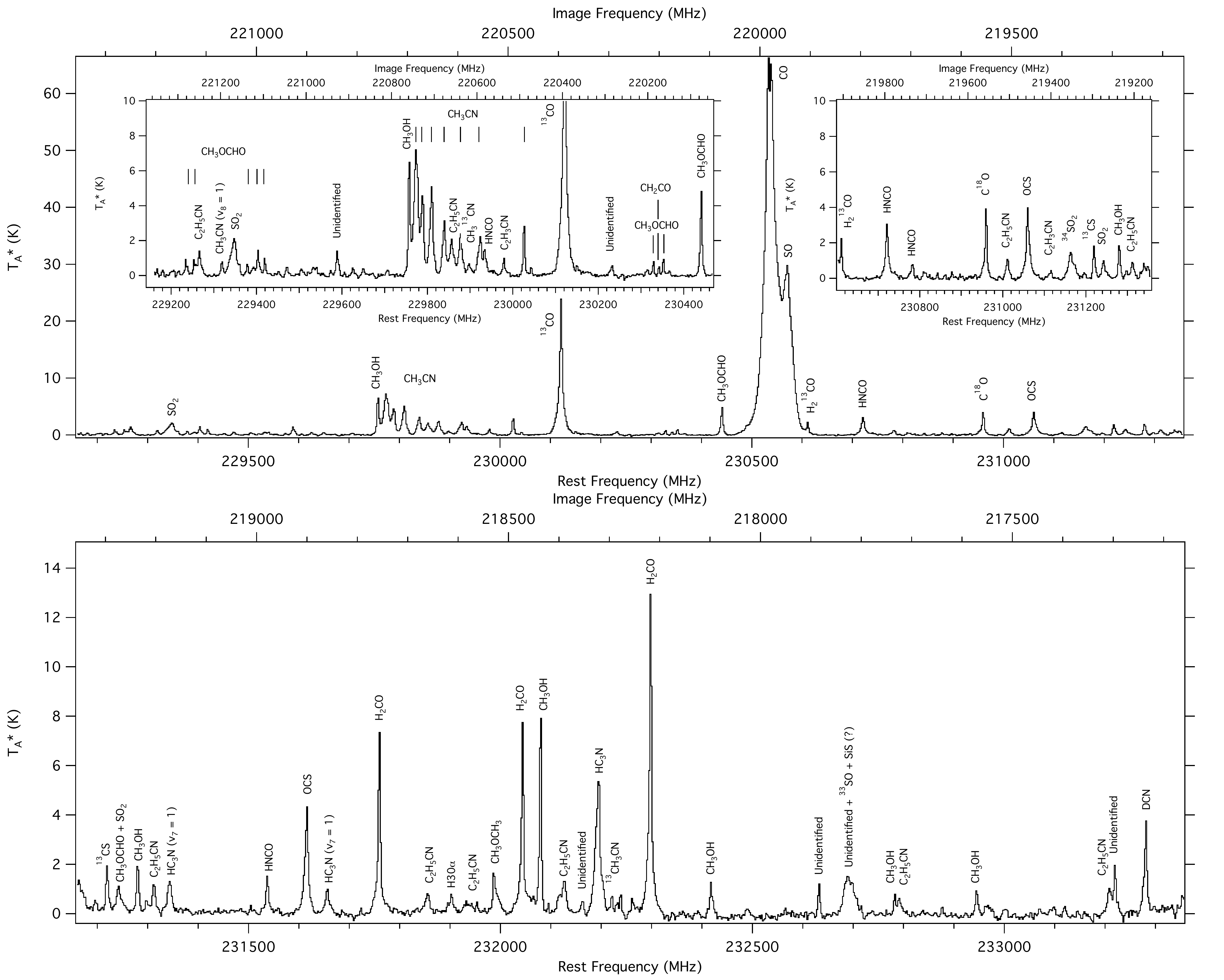}
\caption{\textbf{(Top)} Spectrum of Orion KL from 229160 -- 231356 MHz (Rest) and 219160 -- 221356 MHz (Image).  IF and LO settings were chosen such that CO, $^{13}$CO, and C$^{18}$O are simultaneously acquired in the same spectrometer window.  Insets show detail.  \textbf{(Bottom)} Spectrum of Orion KL from 231160 -- 233356 MHz (Rest) and 217160 -- 219356 MHz (Image).}
\label{ffts2b1}
\end{figure*}

\noindent The target for the final observations was the Orion~KL hot core at (J2000)~$\alpha$~=~05:35:14.5 \break $\delta$~=~-~05:22:31.0 with a $V_{LSR} = 5.0$ km s$^{-1}$.  Pointing was performed on Uranus, and converged to $\sim$1\arcsec.  The CSO 230~GHz double-sideband receiver was used with the Fast Fourier Transform~2 spectrometer providing 4~GHz of bandwidth in each sideband. The typical system temperature was 600~K.  Data were intensity calibrated using the chopper wheel calibration method, with final intensities on the $T_A ^*$ scale.  ON-OFF integrations were collected using a secondary chopping with a throw of 3\arcmin.  Two, 40-second on-source integrations were collected for a total on-source time of 80 seconds.  The data were smoothed to a final resolution of $\sim$2 MHz resulting in an RMS of $\sim$28~mK for line identification (Figure \ref{ffts2b1}).  A first-order baseline was subtracted from the spectra.\\

\noindent A total of 26 species are readily identified in the observations based on comparisons to the \citet{Sutton1985} survey.  Despite the poor weather and short integration time, the CSO detected eight isotopic species, two vibrationally-excited molecules, and twelve complex organic molecules.  More than two dozen additional, weaker features were not assigned.  The reduced data are freely available to the public as Supplementary Information.

\acknowledgments

\end{document}